# Electrochromic chiral ferroelectric nematic liquid crystals

Md Sakhawat Hossain Himel, James T. Gleeson, Robert J. Twieg, Samuel Sprunt, Antal Jákli [*]

Md Sakhawat Hossain Himel and Antal Jákli

Materials Sciences Graduate Program and Advanced Materials and Liquid Crystal Institute, Kent State University, Kent, Ohio 44242, USA

e-mail: mhimel@kent.edu; ajakli@kent.edu

James T. Gleeson, Samuel Sprunt, Antal Jákli

Department of Physics, Kent State University, Kent, Ohio 44242, USA

E-mail: jgleeson@kent.edu; ssprunt@kent.edu, ajakli@kent.edu

Robert J. Twieg

Department of Chemistry and Biochemistry, Kent State University, Kent, Ohio 44242, USA

E-mail: rtwieg@kent.edu

**Conflict of Interest:**

The authors declare no conflict of interest.

**Data Availability Statement:**

The data that support the findings of this study are available from the corresponding author upon reasonable request.

## Abstract

Chiral nematic ($N^*$) liquid crystals are one-dimensional photonic band-gap materials whose reflection wavelength can be well tuned by temperature, but only limited and irreversible tuning can be achieved by electric fields. In contrast, oblique heliconical chiral nematic ($N^*_{ob}$) materials blueshift with <1V/μm fields applied along the helix axis, whereas chiral ferroelectric nematic ($N^*_F$) liquid crystals can be redshifted by <0.1V/μm fields applied perpendicular to the helix axis.

Here we demonstrate that in $N^*_F$ liquid crystals, the reflection color can be reversibly tuned also by electric fields applied along the helix axis. In sandwich cells assembled with bare conducting indium tin oxide (ITO) substrates, the reflectivity peak wavelength increases by up to 200 nm under fields up to 0.4 V/μm. When the ITO substrates are treated with an electrically insulating polymer layer, the reflectivity shift is suppressed. We propose a theoretical model assuming helical deformation of the helix axis under electric field. This model accounts for all observations and also yields an estimate of the $N^*_F$ splay elastic constant which is challenging to determine by other methods. Our findings expand understanding of ferroelectric nematic liquid crystals and suggest potential applications in both tunable reflectors and energy-efficient smart windows.

## I. Introduction

Photonic bandgap materials can both transmit and reflect some part of the visible wavelength of light. Controlling the intensity and wavelength of the reflection has potential applications from energy-saving smart windows [1,2] to augmented reality display devices [3]. Among the best-known one-dimensional photonic bandgap materials capable of selective reflection of light are chiral nematic ($N^*$) liquid crystals (LCs) where there is no spatial order while the average molecular orientation, $\hat{n}$, follows a helix whose axis is perpendicular to $\hat{n}$. [4] The width of the photonic bandgap is $\Delta\lambda = (n_e - n_o)p$, with the gap center at $\lambda = \frac{n_e+n_o}{2} \cdot p \cdot \cos\psi$, where $p$ is the pitch of the helix, $n_e, n_o$ are the extraordinary and ordinary refractive indices, and $\psi$ is the angle between the incident light and the helix axis. [5] Since *p* typically depends on temperature, *N\** LCs have been used as temperature sensors [6,7]. In addition to temperature, the helical structure is also sensitive to external electric and magnetic fields. [8,9] When an electric field is applied perpendicular to the helix axis, $\hat{n}$ is distorted but the helical axis is not. This leads to an increase in $p$ for fields below a critical value, $E_c = \frac{\pi^2}{p}\sqrt{\frac{K_{22}}{\varepsilon_o \Delta \varepsilon}} \sim 10\ V/\mu m$. [8–10] Above this, the helix fully unwinds and the material becomes nematic. The situation is more complicated for electric fields applied parallel to the initial helix axis (for example between ITO electrodes of substrates treated for planar alignment of $\hat{n}$). In a typical case when the bend elastic constant is larger than the twist elastic constant ($K_{33} > K_{22}$), the field induced rotation of the director toward the field must lead to a rotation of the helix axis by an angle $\psi$ with respect to the normally incident light. This will lead to a decrease in pitch and the reflection wavelength, which becomes $\lambda = \lambda_o \cdot cos\psi$. [11–13] As the tilt direction is degenerate, this also leads to an inhomogeneous texture [14,15] and large fields eventually irreversibly destroy the helix. For the hypothetical case when $K_{33} < K_{22}$, it was predicted by Meyer [8] that there is a field-induced transition to a conical helical structure, whereby the helix axis is still parallel to the electric field applied along the helix, but the director turns toward the helix axis. Such a situation is observed experimentally in chiral twist-bend nematic phase [16–19] of flexibly bent-shaped dimers [20,21]. It was found that the reflection band of these materials can be tuned over the entire visible wavelength range in $E \leq 1\ V/\mu m$ electric [22,23] or $B \leq 10\ T$ magnetic [24] fields, so that the reflected color shifts towards shorter wavelengths (blueshift) with increasing fields.

An entirely new class of electrically tunable, selective reflectors was made possible after the discovery of the ferroelectric nematic ($N_F$) liquid crystal phase. [25–30] These materials exhibit spontaneous electrical polarization $\vec{P} \parallel \hat{n}$, where the magnitude of $\vec{P}$ can be as much as $P_s \sim 0.05\ C/m^2$. $N_F$ materials have attracted great interest due to their remarkable physical behavior including formation of freestanding fluid filaments [31–33] and soft piezoelectricity [34].

When the $N_F$ state is made chiral, the resulting $N_F^*$ phase exhibits unique fascinating optical properties, including multiple photonic bandgaps. [35–39] This state has the same structure as $N^*$, with the added element that the permanent polarization follows the same helix as $\hat{n}$. The most relevant result to this study of the $N_F^*$ optics is that in-plane electrical field applied perpendicular to the helical axis can reversibly increase the wavelength of the selective reflection [37,40–42]. The wavelength tuning can be explained as a rotation of the director at the substrates to compensate for the increase of twist elastic energy in the regions where the molecular orientation changes rapidly. The reflectivity tuning is due to the distortion of the helix that exists even when the director cannot rotate at the substrates. [40] Although windows utilizing such in-plane reflectivity tuning can be used for low power temperature control [43], they use interdigitated electrodes that can cause light diffraction and they require relatively large voltages for operation.

Here we report studies on films in $N_F^*$ phase when the electric field is applied uniformly along the helix axis. In stark contrast to prior observations on $N^*$ films, when the material in $N_F^*$ phase is sandwiched between untreated, electrically conductive substrates, we find a reversible increase in the reflectivity peak up to $200 nm$ with applied field less than 0.4V/μm. When the same material is prepared using an insulating polymer coating on the electrodes, we observe an overall decrease in reflectivity as the applied field is increased, but no shift in peak wavelength. We propose a simple model to explains the electric field dependence of the peak wavelength shift on electrically conducting substrates, and the suppression of the tuning in presence of insulating polymer layers. Additionally, it gives an estimate of the splay elastic constant that is challenging to measure by other experiments.

## II. Materials and methods

For our studies we chose a mixture, designated KPA-02 doped with a chiral additive, R-5011, with estimated helical twisting power of $HTP \sim 120 \mu m^{-1}$. [40] KPA-02 contains 60 wt% of the single component ferroelectric nematic liquid crystals compound RT12155 [4-[(4-Cyano-3,5-difluorophenoxy)carbonyl]phenyl 2-n-propoxybenzoate] (see Figure 1(a)) and 40 wt% of a commercially available nematic LC HTG-135200-100 (clearing point $T_c = 97°C$) purchased from HCCH. On cooling from the isotropic phase, KPA-02 transitions to the $N_F$ phase at 47°C and remains stable at room temperature. The spontaneous polarization of KPA-02 is measured to be $P_s \approx 0.044 \frac{C}{m^2}$. [40] The chiral dopant R-5011 (see Figure 1(a)) was added to KPA-02 at $c \approx$ 3.0 and 3.4 $wt\%$ concentrations. This material has $N_F^*$ phase at ambient temperature and exhibits tunability in both the reflected color as well as overall reflectivity under in-plane DC field. With AC field (100Hz) only the overall reflectivity is tunable. [40]

We prepared sample cells using 0.7 mm transparent glass plates with $\sim 10\ nm$ ITO conductive coatings. A potential difference applied between the ITO on opposing plates provides the electric field. We note that the surface anchoring energy on untreated ITO is usually weak ($W \leq 10^{-6} J/m^2$) [44]. This means both polar and azimuthal angles between the director and the surface are not rigidly fixed. In addition, we prepared samples in which the ITO layer was coated with $\sim 50 nm$ thick polyimide, PI2555. This is a widely used alignment layer that provides $W_p \sim 10^{-4}$-$10^{-3} J/m^2$ polar (and typically 10x weaker azimuthal) anchoring strength for liquid crystals. [45]

Figure 1(b) shows the experimental setup to measure the voltage dependent reflectance. An HP 33120A function generator and an FLC F20AD voltage amplifier were used to apply the electric fields. The wavelength dependent reflectivity $R(\lambda)$ was measured using an OceanOptics VIS-IR spectrophotometer.

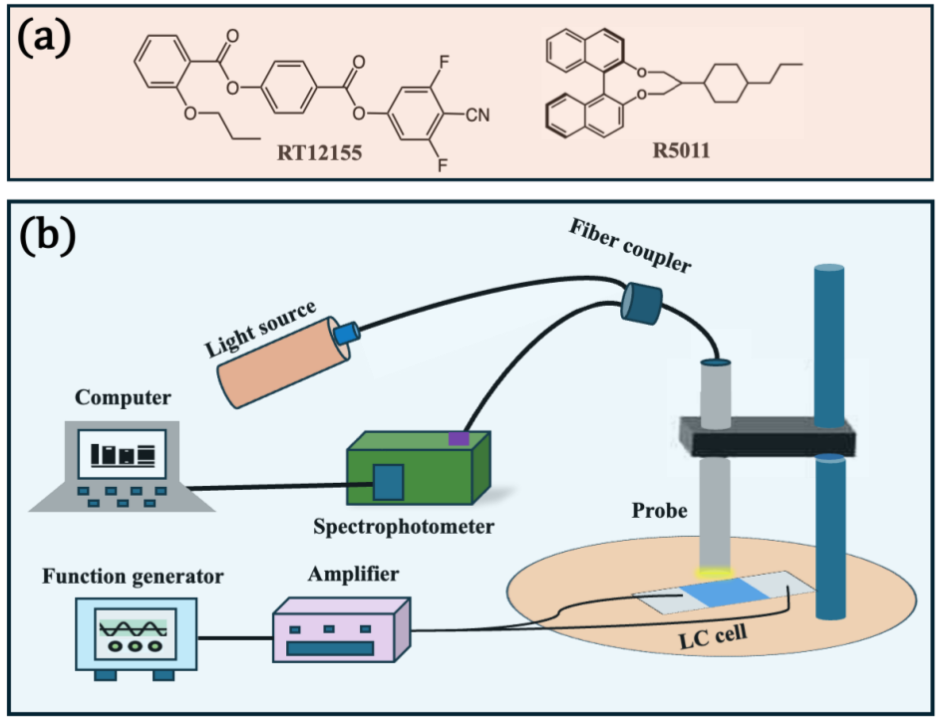

*Figure 1: Materials and Methods. (a) Molecular structure of the main component (RT12155) of the FNLC mixture KPA-02 and of the chiral dopant R 5011. (b) Experimental setup to measure the wavelength dependence of reflectivity.*

## III. Results

The reflectivity spectra $R(\lambda)$ of an $L = 30$ μm thick KPA-02 +3.0% R5011 sample under various 100 Hz electric fields applied between bare ITO covered substrates are shown in Figure 2(a). In contrast to apolar, $N^*$ materials, the selective reflection is sensitive to electric fields well below $1 V/\mu m$ fields. The peak of $R(\lambda)$ increases from 480 nm at zero field to 670 nm at 0.4 $V/\mu m$.

The magnitude of the reflectivity only slightly varies between 470 nm and 570 nm and then decreases monotonously and reaches zero at $\lambda \sim 700\ nm$. The electric field dependence of the peak of the selective reflection wavelength is shown in Figure 2(b). The field-dependence can be fitted with reasonable accuracy to a quadratic function, $\lambda\ (nm) = 480 + 970 \cdot E(V/\mu m)^2$. The appearance of the cell at different fields under white light reflection microscopy is shown in the inset to Figure 2(b). One can see various uniformly colored textures that show no trace of the electro-hydrodynamical patterns seen in the case of conventional $N^*$ materials. [12,14] The observation that the maximum reflected intensity is less than 15%, indicates imperfect planar alignment.

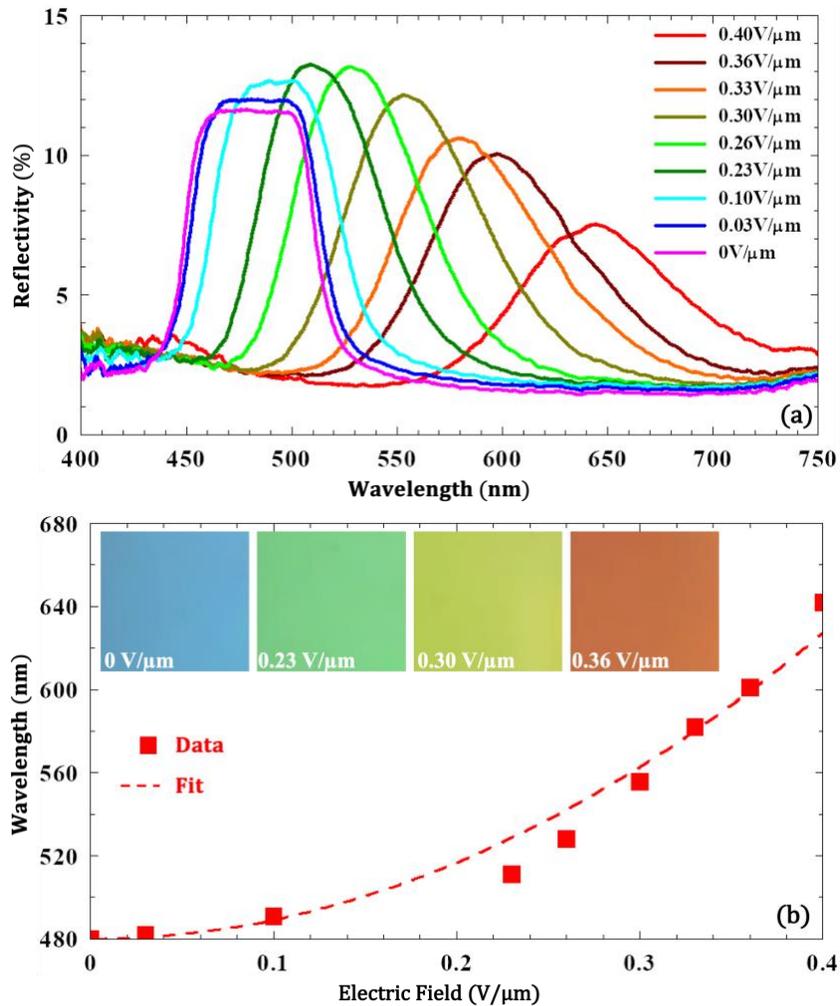

*Figure 2: Summary of the reflectivity of a 30 μm KPA-02 doped with 3 wt% R5011 at room temperature under various 100 Hz electric fields applied between ITO coated substrates without any alignment layer. (a): Wavelength dependent reflection spectra; (b): Electric field dependence of the peak selective reflection wavelength. Dashed line is a fit to Eq.(5) in the text. Insets show $\sim 120\ \mu m\ \times\ 100\ \mu m$ size microscopic images in reflection under various electric fields.*

The time dependence of the transmitted white light intensity is illustrated in Figure 3 when 100 $Hz$, 40 $V$ square wave packets are turned OFF and ON. In contrast to the samples driven with in-plane fields applied perpendicular to the helix axis [40], in the present work where the field is applied along the helix axis, the reflectivity color and intensity recover quickly after a very short transient milky scattering state.

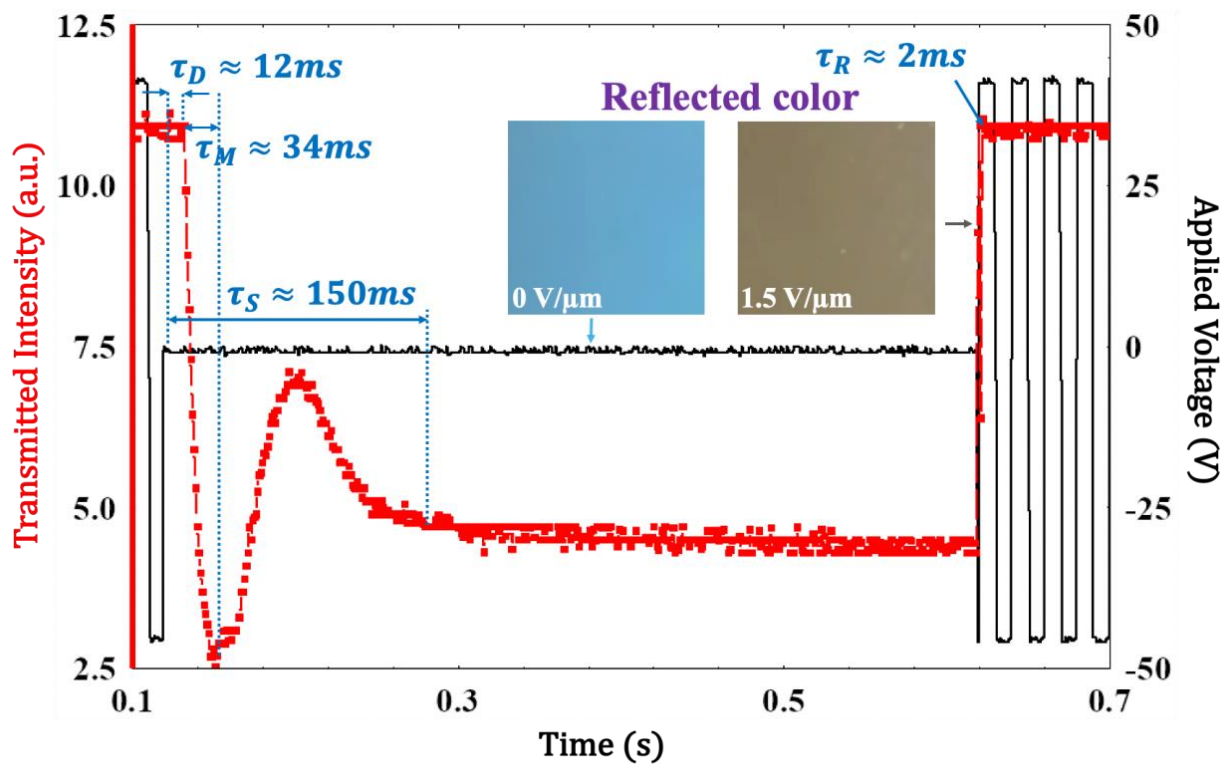

*Figure 3: Time dependence of the transmitted intensity observed in a 30 μm cell of KPA-02 + 3% R5011 mixture at room temperature. The red curve plotted against the left axis is the transmitted white light intensity, and black line shows the applied packets of the 100 Hz square-wave voltage plotted against the right axis.*

It can be seen that after a $\tau_D \approx 12\ ms$ delay, the transmittance decreases quickly and reaches a minimum after $\tau_M \approx 34\ ms$. This is the stage where the sample is scattering. The transmittance with the original blue color reaches its steady state after $\tau_S \approx 150\ ms$. Remarkably, the high transmittance (dark in reflection) state is reached quickly, within $\tau_R \approx 2\ ms$ rise time. The reflected colors in the OFF and ON states are shown in the inset images.

A summary of the results on 10 $\mu m$ thick KPA-02+3.4 % R 5011 planar aligned cells with different alignment layers is presented in Figure 4. Figure 4(a) shows the wavelength-dependent spectra at various fields for a sample between bare ITO substrates under increasing (solid lines) and decreasing (dotted lines) fields. One can see that the field dependence of the reflected wavelength is fairly well reproducible in the increasing and decreasing ramps. As the concentration of the chiral dopant is ~15% larger than for the 30 µm cell shown in Figure 2, the zero-field reflection color has ~15% lower wavelength (~415 nm) than for the 30 µm film with 3.0 % chiral dopant. The maximum reflectivity is again less than 20 %, indicating imperfect planar alignment. Similar to the 30 µm film between bare ITO coated plates, the reflection redshifts by ~200 nm between 0 and 0.4 V/µm applied fields.

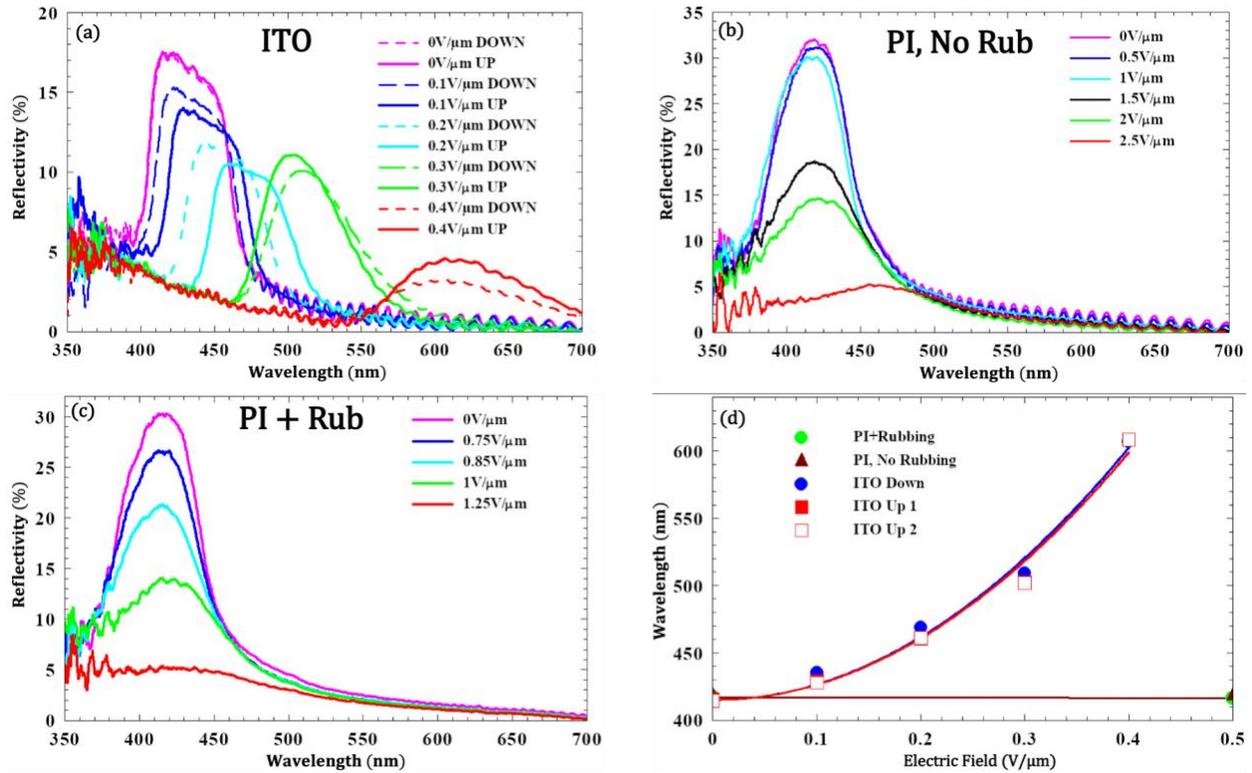

*Figure 4: Wavelength and electric field dependences of reflectivity on 10 µm thick KPA-02+3.4 % R 5011 planar aligned cells with different alignment layers. (a): Sample between bare ITO substrates (solid/dashed lines measured in increasing/decreasing fields) ; (b): Sample between non-rubbed PI2555 layers;(c) Sample between uniformly rubbed PI2555 layers; (d) 100 Hz electric field dependence of the positions of the maxima of the reflected wavelengths for samples with different substrates. Solid lines represent fits to Eq.(5) in the text.*

The R($\lambda$) curves measured in a 10 µm sample cell with unrubbed PI 2555 polyimide alignment layers are shown in Figure 4(b). Similar to the cell with untreated ITO plates, the zero field reflection peak wavelength is also ~415 $nm$, however, the maximum reflectivity at zero field is now over 30%, indicating much more uniform planar alignment. Additionally, the reflection peak does not change below about $2V/\mu m$ field, and the reflected intensity is also much less dependent on the electric field than the sample between bare ITO layers. Results on samples contained between uniformly rubbed PI2555 (Figure 4(c)) are very similar to those for the case of unrubbed PI2555 (similar reflectivity and similar field dependences).

Figure 4(d) compares 100 $Hz$ electric field dependences of the maxima of the reflected wavelength for different runs on the uncoated ITO substrates and on the PI2555 alignment layers. Just as for the 30 µ$m$ film with 3.0 $wt\%$ chiral dopant shown in Figure 2, the peak reflectivity wavelength depends quadratically ($\lambda = 415 + 1170 \cdot E^2$) on the electric field for the 10 µm cell with bare ITO substrates. However, with PI2555 coated substrates, in the 0-0.5 V/µm range the reflectivity peak wavelength does not change with field.

The results of the KPA-02 + 3.4 wt% R5011 samples between rubbed polyimide substrates at 5, 20 and 30 µm film thicknesses are compared in Figure 5.

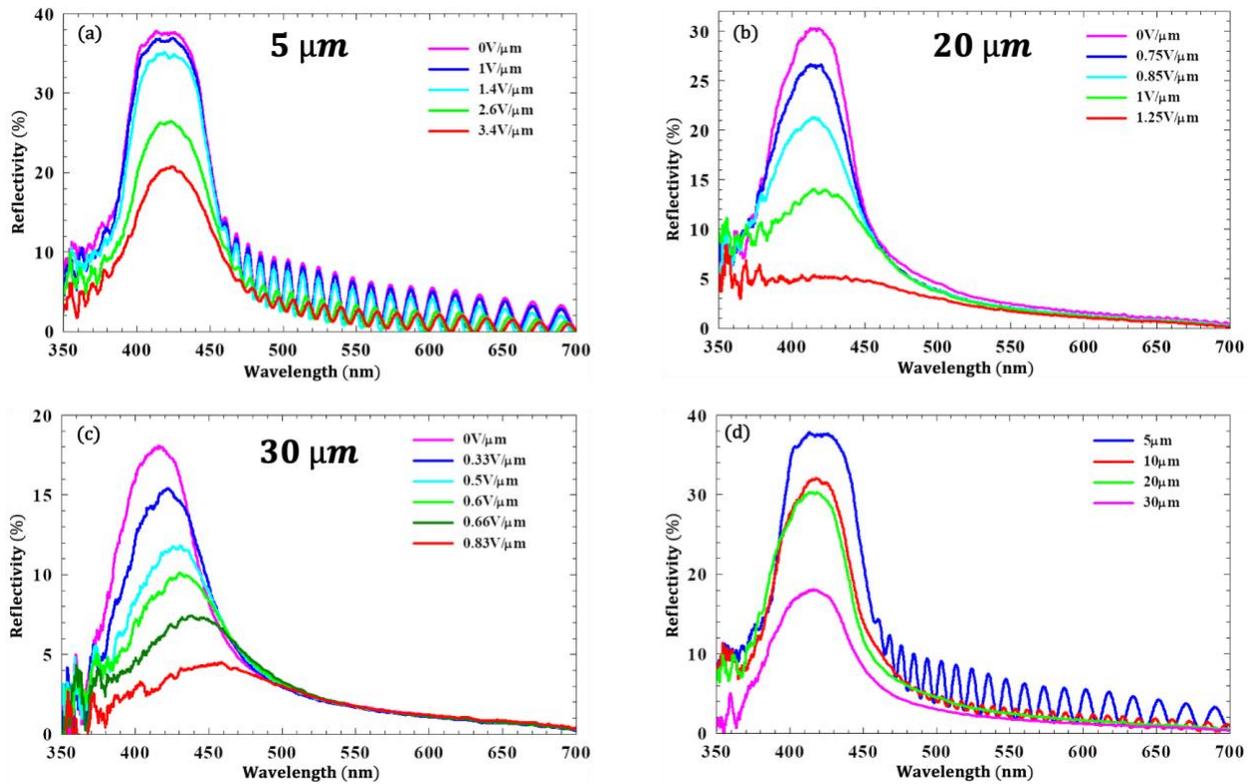

*Figure 5: Reflection spectra of KPA-02+3.4 wt% R5011 samples between rubbed polyimide substrates. (a): 5 µm; (b): 20 µm; (c): 30 µm films under various electric fields. (d): Comparison of zero field spectra for 4 different film thicknesses.*

Figure 5(a) shows that the reflectivity peak wavelength of a 5 μm film does not change between 0 and 3.4 $V/\mu m$ field but only that the overall reflectivity decreases from ~38% to ~20%. As seen in Figure 5(b) on a 20 μm film, the peak wavelength shifts from 416 $nm$ to 422 nm between 0 and 1 V/μm, and the overall reflectivity decreases from 30 % at 0 $V$ to less than 5% at 1.25 $V/\mu m$. The 30 μm cell (see Figure 5(c)) shows a larger change in the peak reflectivity wavelength from 416 $nm$ at 0 $V/\mu m$ to 425 $nm$ at 0.4 $V/\mu m$ and to 455 nm at 0.83 V/μm while the reflectivity decreases from 18% to 4%. However, this still represents a more than 20-fold decrease compared to the 30 μm cell between bare ITO substrates. Figure 5(d) compares the zero-field reflectivity for 5, 10, 20 and 30 μm thick films. One can see that while the reflection color is independent of the film thickness for the same chiral dopant concentration, the reflectivity decreases from 38% to 18% when the film thickness increases from 5 μm to 30 μm, indicating less uniform alignment as the film thickness increases.

## IV. Discussion

Unlike ordinary apolar chiral nematic liquid crystals, the chiral ferroelectric nematic liquid crystal mixture studied here exhibits reversible response to electric fields applied along the zero-field helical axis. The response is the most pronounced in films contained between uncoated ITO substrates. In this case the peak reflectivity wavelength increases by ~200 nm and the overall reflectivity decreases by about 50% for fields up to $0.4\frac{V}{\mu m}$ (on both 10 μm and 30 μm thick samples). The change in R(λ) was found to be reversible (see Figure 4(a)) with no observation of large-scale director deformation. Samples in cells prepared with PI2555 coating on the ITO showed only very small (if any) change in peak reflectivity wavelength and smaller change in overall reflectivity. Films with thickness ≤ 10 μm showed no change in peak reflectivity wavelength below $2\frac{V}{\mu m}$, while the 30 μm film had about a 20-fold weaker field dependence than that of the 30 μm film between bare ITO. Interestingly, the rubbing conditions of the PI2555 did not affect the results.

Below we present a simple model that accounts for these observations. The basic premise is the linear coupling between $\vec{E}$ and $\vec{P}$ very strongly promotes decreasing the angle between them. One manner in which that might happen would be for the helical state at zero field to become heliconic (so that the director is not strictly perpendicular to the helix axis). Even if $K_{33} < K_{22}$ were true as for $N^*_{ob}$ [22], we would expect to see a decrease of the pitch and therefore of the peak reflectance wavelength. As this is not the case, we therefore consider a distortion of the helical axis caused by the polar coupling.

Our starting point is the free energy density for chiral nematic materials, assuming negligible surface anchoring. As the sketch in Figure 6 shows, in both the $N^*$ and the $N^*_F$ phases, the director deformations are mainly twist and splay and the free energy density can be approximated by

$$f = \frac{1}{2}K_{22}[\hat{n}\cdot(\nabla\times\hat{n}) - q_o]^2 + \frac{1}{2}K_{11}(\nabla\cdot\hat{n})^2 - \frac{1}{2}\varepsilon_0\Delta\varepsilon E^2 - \vec{P}_S\cdot\vec{E}. \qquad (1)$$

Here $K_{22}$ and $K_{11}$ are the twist and splay Frank elastic constants, $\Delta\varepsilon$ is the dielectric anisotropy, $P_S$ is the ferroelectric polarization and $q_o = \frac{2\pi}{p_o}$ is the zero field wavenumber of the helix with pitch $p_o$.

In the apolar $N^*$ phase [see Figure 6(a)], the up and down director tilt is energetically equivalent due to the dielectric coupling, so the twist elastic energy term of Eq.(1) is minimized when the pitch does not change. For the same reason, the helical axis can tilt uniformly by an angle $\psi$ with respect to the applied field, therefore the splay elastic term is zero. Consequently, in $N^*$ the reflection wavelength decreases as the helix axis tilts according to $\lambda = \frac{n_o+n_e}{2}\cdot p\cdot\cos\psi$. We note that any tilt direction normal to the electric field is equivalent. This leads to inhomogeneous director structure as was indeed observed experimentally. [11]

By contrast, in $N^*_F$ materials, the linear ferroelectric coupling dominates, therefore, there is a difference whether the polarization tilts towards or away from the field. In order that the polarization $P_S$ can everywhere turn toward the field direction, the shape of the helix axis has to be deformed periodically with pitch $p = p(E)$ so the polarization rotate in the opposite sense at positions separated by $p/2$ along the helical axis. This deformation can be 2-dimensional or 3-dimensional. As the directions normal to the electric field are equivalent, such as in $N^*$ films, one would expect formation of differently aligned domains under electric fields. As the texture remained uniform under electric field applied along the undistorted (zero field) helix axis, we assume the deformation is 3-dimensional with cylindrical symmetry. One such deformation of the helix axis that would also reflect the chirality of the material would be a helical winding or coiling of the axis itself. In such structure the polarization direction would uniformly make an angle $90° - \psi_o$ with the applied field as illustrated in Figure 6(b). We emphasize that this structure is different from the oblique heliconical $N^*_{ob}$ phase where the director has heliconical distribution with respect to the straight helix axis. In our case the polar director is everywhere perpendicular to the coiling helix axis.

For such a deformed $N^*_F$ structure the polarization has a non-zero splay deformation with a positive free-energy contribution proportional to $(q \cdot sin\psi_o)^2$, where $q = \frac{2\pi}{p}$. This splay elastic energy density term decreases as $p$ increases, thus driving the equilibrium toward increasing pitch under increasing electric fields. This leads to an increase of the peak reflection wavelength as $\lambda = \frac{n_o+n_e}{2} \cdot p(E) \cdot cos\psi_o$.

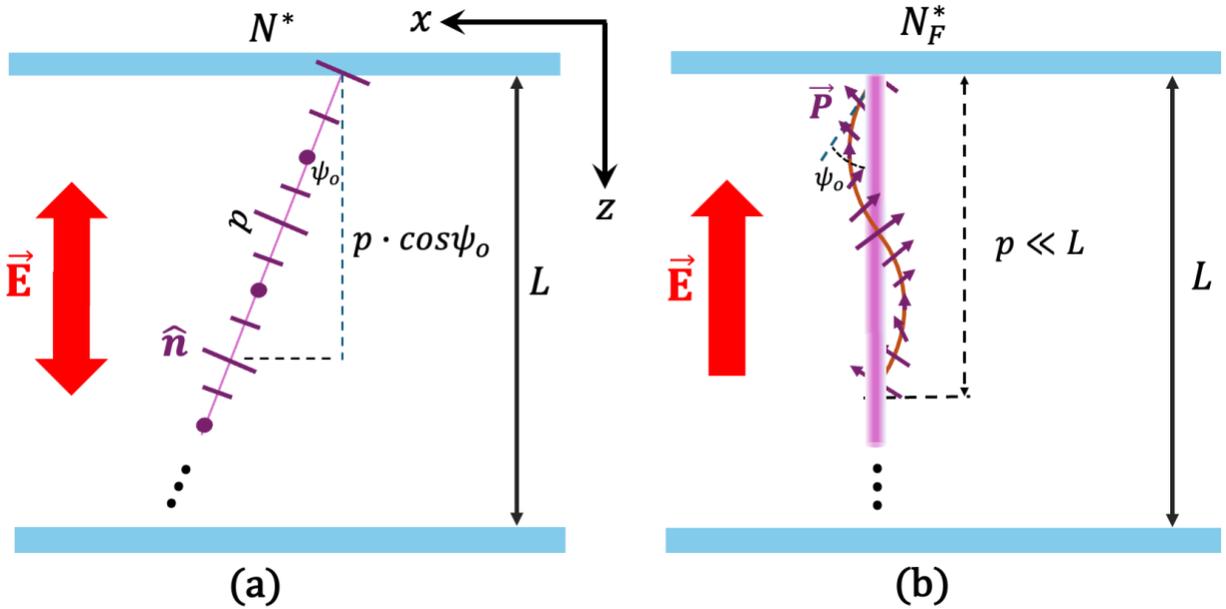

*Figure 6: Illustration of the effect of an external electric field applied along the helix axis in the $N^*$ phase (a), and in $N^*_F$ phase (b). Purple lines (arrows) indicate the direction of the director in the $N^*$, and the ferroelectric polarization in the $N^*_F$ phases.*

Neglecting the dielectric term compared to the ferroelectric coupling, Eq. (1) can be rewritten as

$$f = \frac{1}{2}K_{22}(q - q_o)^2 + \frac{2\pi^2 K_{11}}{p^2} sin^2\psi_o - P_S \cdot E \cdot sin\psi_o. \qquad (2)$$

Assuming small helix axis deformation, $sin\psi_o \approx \psi_o$, the free energy per unit area $G = \int_o^L f(z)dz = f \cdot L$ can be written as

$$G \approx L[2\pi^2 K_{22}(\frac{p-p_o}{p_o p})^2 + \frac{2\pi^2 K_{11}\psi_o^2}{p^2} - P_S E \psi_o]. \tag{3}$$

The field dependence of the equilibrium director tilt $\psi_o$ with respect to the original planar direction at the boundary can be obtained from the $\frac{\partial G}{\partial \psi_o} = 0$ condition, giving

$$\psi_o \approx \frac{P_S E p^2}{4\pi^2 K_{11}}. \tag{4}$$

Substituting this back to Eq.(3), we get the electric field dependence of the wavenumber from the $\frac{\partial G}{\partial p} = 0$ condition. This gives for $\Delta p = p - p_o \ll p_o$ that

$$\Delta p \approx \frac{P_S^2 \cdot p_o \cdot p^4}{16\pi^4 \cdot K_{22} \cdot K_{11}} E^2. \tag{5}$$

This implies that at low fields that shift of the pitch $\Delta p \ll p_o$ increases as $E^2$ in agreement with our experimental observations shown in Figure 2(b) and Figure 4(d). The deviation from this behavior at higher fields is due to $\Delta p$ is becoming comparable to $p_o$.

Assuming the extraordinary (ordinary) refractive index is $n_e \approx 1.7$ ($n_o \approx 1.5$), we get the average refractive index $\bar{n} \approx 1.6$ and the wavelength of the selective reflection $\lambda = \bar{n} \cdot p \approx 480\ nm$ for the mixture with 3.0% R5011. This means that the pitch increases from $p_o \approx 300\ nm$ to $p(0.3\ V/\mu m) \approx 350\ nm$, i.e., $\Delta p\left(0.3\frac{V}{\mu m}\right) \approx 80 nm$. With these parameters we estimate $K_{22} \cdot K_{11} \approx 4 \cdot 10^{-21}\ N^2$. Assuming that $K_{22} \sim 10\ pN$, this implies a very large value for $K_{11} \sim 400\ pN$. Using these estimates, Eq.(4) predicts the tilt of the helix axis is $\psi_o \approx 6°$ at $E = 0.3\ V/\mu m$. This validates the small angle assumption. We note our estimated value of $K_{11}$, is about 40 times larger than the splay elastic constant typical for apolar nematic liquid crystals. However, this is not surprising as in $N_F$ materials the splay elastic constant is expected to be up to two orders of magnitude larger than in normal nematic materials. [46] This is because splay, which is a non-zero divergence in $\vec{n}$, induces a bound charge density: $-\vec{\nabla} \cdot \vec{P} = \rho$ that leads to a large electrostatic energy. [46–48]

To explain the suppression of the peak wavelength shift for cells with PI2555 coating (see Figure 4(b-d) and Figure 5), first we consider the effect of the strong polar anchoring that will add a term $W_p \cdot sin^2\psi_o \approx W_p \cdot \psi_o^2$ [49] to $G$ at the boundaries, where $W_p$ is the polar surface anchoring energy per unit area. Comparing this to the splay elastic term in Eq.(3), which has the same angle dependence, we get that $\frac{2\pi^2 L K_{11} \psi_o^2}{p^2} \approx 2 \cdot \psi_o^2 \gg W_p \cdot \psi_o^2$ even for $W_p$ as large as $\sim 10^{-3} J/m^2$. This shows that the anchoring strength has only a minor effect on the tuning of the reflection wavelength.

Next, we consider the effect of the insulating nature of the polyimide coating, which leads to a depolarization field $E_{dep} = -\frac{P_s \cdot sin\psi_o}{\varepsilon_o \cdot \varepsilon}$ due to bound charges ending on the insulating surface. Adding this term to the applied field $E$ in Eq. (2), we get a term $\frac{P_s^2 \cdot sin^2\psi_o}{\varepsilon_o \cdot \varepsilon}$ that will rescale the splay elastic term in (2) so that the effective splay elastic coefficient $\widetilde{K}_{11}$ becomes

$$\widetilde{K}_{11} = K_{11} + \frac{p^2 P_s^2}{2\pi^2 \varepsilon_o \cdot \varepsilon} \tag{6}$$

The apparent dielectric constant $\varepsilon$ in (6) was shown in N$_F$ phase to be related to capacitance of the insulating layer as $\varepsilon = \varepsilon_I L/L_I$. [50,51] Here $\varepsilon_I \approx 3.5$ is the dielectric constant of the polyimide PI2555 [52]; $L_I$ is the sum of the thicknesses of the insulating layers on the two substrates and $L$ is the thickness of the liquid crystal film. In our experiments $L_I \approx 100\ nm$, at $E = 0.4\ V/\mu m\ p \approx 0.4\ \mu m$ and $P_s \approx 0.04\ C/m^2$ which give that $\widetilde{K}_{11} \sim K_{11} + \frac{5 \cdot 10^{-8} N}{L\ (\mu m)}$. For a 5 μm cell well below the relaxation frequency this means $\widetilde{K}_{11} \sim K_{11} + 10\ nN \sim 28 K_{11}$, while for a 30 μm film it decreases to $\sim 5 K_{11}$. This qualitatively explains the observation that the electric field induced wavelength tuning in between PI2555 is >25 and >5 times smaller, respectively in the 5 μm and on the 30 μm films bounded by PI2555 boundary layers on top of ITO compared to the samples bounded by bare ITO layers (see Figure 5).

## V. Conclusions

We have presented the first observations of the reflectivity for an electric field applied along the average helix axis of chiral ferroelectric nematic ($N_F^*$) liquid crystals with helical pitch in the range of $p_o \sim 0.3\ \mu m$. In stark contrast to observations on apolar $N^*$ [11], we observed a large increase in peak reflectivity wavelength approximately proportional to the square of the electric field when the film is contained between uncoated ITO substrates. Such an increase was found to be reversible up to $E \leq \frac{1V}{\mu m}$ without leading to an inhomogeneous pattern. Alongside to the increase of the peak wavelength, the intensity of the reflection was decreasing in increasing fields. Such results are significant as the field can be applied between the ITO electrodes thus requiring smaller voltages than the field perpendicular to the helix axis. This configuration also offers high resolution pixelation for display applications or can provide uniform clear state for smart window functions.

Measurements on films sandwiched between insulating PI2555 polyimide layers coated on the ITO electrodes showed more than an order of magnitude suppression of both the peak wavelength shift and on the intensity of the reflection.

We proposed a simple theory that explains all details of our experimental results and even provides an estimate of the splay elastic term which is about 40 times larger than in dielectric nematic materials; a result that is in qualitative agreement with recent results on another $N_F$ material. [48]

The observed tuning of the reflection wavelength and intensity happens at low fields and does not require alignment layers and patterning in-plane electrodes. For these reasons, they offer applications as tunable reflectors [40] and low power temperature control [43] that so far have been demonstrated in $N_F^*$ materials with in-plane electrodes only.

## VI. Acknowledgement

This work was supported by US National Science Foundation grant DMR-2210083.